\documentclass[aps,prb,twocolumn,groupedaddress,floatfix]{revtex4}
\usepackage{graphicx}
\usepackage{dcolumn}
\usepackage{bm}
\usepackage{amsfonts}
\usepackage{amssymb}
\usepackage{amsmath}
\usepackage{graphicx}
\usepackage{bm}

\begin{document}
\title{Inelastic carrier lifetime in a coupled graphene / electron-phonon system: Role of plasmon-phonon coupling}


\author{Seongjin Ahn$^{1}$}
\author{E. H. Hwang$^{2}$}
\email{euyheon@skku.edu}
\author{Hongki Min$^{1}$}
\email{hmin@snu.ac.kr}
\affiliation{$^1$ Department of Physics and Astronomy and Center for Theoretical Physics, Seoul National University, Seoul 151-747, Korea}
\affiliation{$^2$ SKKU Advanced Institute of Nanotechnology and Department of Physics, Sungkyunkwan University, Suwon, 440-746, Korea}

\date{\today}
\begin{abstract}
We calculate the inelastic scattering rates and the hot electron
inelastic mean free paths for both  monolayer and bilayer graphene on
a polar substrate. We study the quasiparticle self-energy by taking
into account both electron-electron and electron-surface-optical-phonon interactions. 
In this calculation the leading-order dynamic
screening approximation (G$_0$W approximation) is used to obtain 
the quasiparticle self-energy by treating electrons and phonons on an equal footing.  
We find that the strong coupling between the surface-optical phonon and the plasmon leads to an
additional decay channel for the quasiparticle through the emission of the
coupled mode, and gives rise to an abrupt increase in the scattering
rate, which is absent in the uncoupled system. 
In monolayer graphene a single jump in the scattering rate occurs,
arising from the emission of the low-energy branch of the coupled plasmon-phonon
modes. In bilayer graphene the emission of both low- and high-energy
branches of the coupled modes contributes to the scattering rate and gives
rise to two abrupt changes in the scattering rate. The jumps
in the scattering rate can potentially be used in hot electron
devices such as switching devices and oscillators. 

\end{abstract}

\maketitle


\section{Introduction}

Two-dimensional (2D) graphene has been extensively studied during
recent years because of its fundamental and technological interest. \cite{sarmarmp2011,netormp2009} 
Although it is possible to grow graphene on non-polar substrates,\cite{Wu2011} in most currently available graphene samples (e.g., graphene field-effect transistors) graphene lies on top of a polar substrate such as SiO$_2$,\cite{chennatnano2008, Dorgan2010, Fei2011} SiC,\cite{Robinson2009, Sutter2009, Liu2010, Koch2010} hBN\cite{dean2010,  wang2013, Ulstrup, Principi} or HfO$_2$.\cite{Zou2010}
In such graphene samples the polar optical phonons of the substrate
are localized near the graphene-substrate interface, and free carriers
in graphene couple to the surface polar (SO) phonons of the underlying
substrate through the long-range polar Fr\"{o}hlich coupling. In polar materials 
the longitudinal optical (LO) phonons generate a long-range electric field
which scatters electrons, and typically their contribution to
resistivity is dominant at room temperatures. \cite{andormp1982} 
In non-polar materials such as graphene, however, the non-polar optical phonons
have little effect on carrier transport because the long-range interaction
between electrons and phonons is absent and the energy of the (non-polar) optical phonon is 
very high, $\sim$200 meV. 
In addition, contributions from acoustic phonons\cite{hwang2008,park2014, sohier2014, Zhang2014} are relatively small because the electron-acoustic-phonon coupling is rather weak in graphene due to the small deformation potential.\cite{min2011, efetov2010, Pachoud2010, dawlaty} For this reason, the SO phonons rather than
non-polar LO or acoustic phonons can be the dominant scattering
source at room temperature in graphene on a polar substrate. \cite{chennatnano2008,Zou2010,Yan2013}




In addition to being the main scattering source in transport, it is
well known that in polar materials the electron-polar-optical-phonon 
interaction leads to polaronic many-body renormalization of the single
particle properties, e.g., polaronic Fermi velocity
(effective mass) renormalization and broadening of the
quasiparticle spectral function.\cite{Mason1985,mahan, Tse2007,LeBlanc2011,Carbotte2013} 
Even though these effects are
expected to be small in graphene due to the weak Fr\"{o}hlich coupling
arising from the spatial separation between electrons in
graphene and the surface of a substrate and due to the large
dielectric constant of the substrate, there is a much stronger quantitative
manifestation of electron-SO-phonon coupling in graphene on a
polar substrate, which is the macroscopic coupling of the electronic collective
modes (i.e., plasmons) to the SO phonons of the system via the
long-range Fr\"{o}hlich coupling.  This mode coupling phenomenon, which
hybridizes the collective plasmon modes of the electron gas with the
SO-phonon modes of the substrate, gives rise to coupled
plasmon-phonon modes, which have been extensively studied both
experimentally\cite {Liu2010, Zhu2014, Fei2011,Koch2010, Brar2014} and theoretically\cite{ph_el, Jablan2011, Ong2012} in graphene. The
plasmon-phonon coupling is important in many single-particle
properties including the inelastic carrier life time, hot-electron
energy-loss processes, and transport properties.

The objective of this paper is to provide the inelastic carrier lifetime and the
inelastic mean free path of monolayer and bilayer graphene in the presence of the long-range polar  
Fr\"{o}hlich coupling between electrons in graphene and SO phonons in
the underlying polar substrate within the leading-order perturbation theory, i.e., G$_0$W
approximation. We consider the effective total interaction (i.e., the
coupled electron-electron and electron-SO-phonon interaction) 
within the framework of the random phase approximation
(RPA). Even though the problem is treated within the G$_0$W framework of the
leading-order effective interaction approximation, our results should
be quite valid in graphene because graphene has a very weak
Fr\"{o}hlich coupling, which justifies the neglect of the electron-phonon
vertex corrections. We include, however, important physical
effects of the dynamical screening, phonon self-energy correction,
plasmon-phonon mode coupling, and Landau damping. 
In the presence of electron-SO-phonon coupling we find added
features in the inelastic carrier lifetime, which are obviously absent
without the coupling.

There have been several studies upon the effects of electron-SO-phonon
interaction in graphene.\cite{SSC,scharf2013,fratini2008,Schiefele2008,Li2008} Various
many-body quantities such as- the self-energy, scattering rate, and spectral
function have been calculated for the statically screened electron-SO-
phonon interaction. Such calculations based on the static screening
approximation are justified when the charge carrier density is high
enough that the corresponding Fermi energy exceeds the SO phonon
energy. However, the SO phonon energies of the common polar substrates
for graphene such as SiC or SiO$_2$ range from 50 to 200 meV,
and they are comparable to the typical Fermi energies of graphene, 100 --
300 meV. Thus the dynamic screening of the electron-SO-
phonon interaction is more desirable. \cite{Ong2012}	
In addition, because the SO phonon energy
and the energy of electrons are comparable, 
we cannot treat the electron-electron interaction and the electron-SO
phonon interaction separately. Therefore, we
need to treat both interactions equivalently within the same
dynamic screening approximation. Quasiparticle properties of various
systems have been calculated considering both electron-electron and 
electron-phonon interactions.\cite{2DEG, hwang1995} However, there appears to be a lack of calculation
of quasiparticle properties of graphene on a polar substrate with both
electron-electron and electron-SO-phonon interactions equally
treated.  

In this paper, we calculate the scattering rate ($\tau$$^{-1}$) and
the corresponding inelastic mean free path ($l$) of quasiparticles in both
monolayer and bilayer graphene by taking into account both 
electron-electron and electron-SO-phonon interactions.
 We also investigate the effect of the dynamic
screening of the electron-SO-phonon interaction and plasmon-phonon
coupling. In addition, we propose a possible technological application
to a lateral hot-electron transistor by making use of the electron-SO-
phonon interaction effect. Throughout the paper, we refer to the
graphene system with both electron-electron and electron-SO-phonon interactions as the coupled system,
whereas the graphene system with only the electron-electron interaction as
the uncoupled system. 

The paper is organized as follows. In Section II the generalized theory
is presented for calculating the electron self-energy in the presence
of both electron-electron and electron-SO phonon interactions. Section
III presents the results for the scattering rates and the hot electron
mean free paths for both monolayer and bilayer graphene. We summarize
in Sec. IV with a discussion. 

\begin{figure}

\centering
	\includegraphics[scale=0.28]{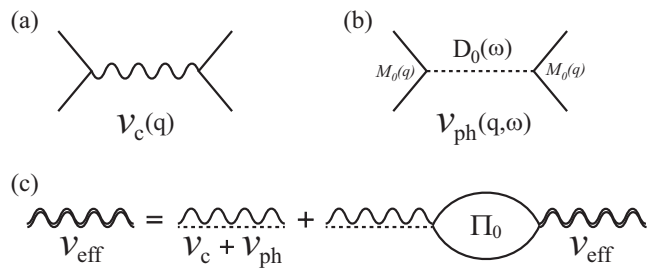}		
\caption{\label{fig:wide} (a) Electron-electron Coulomb interaction. (b) Phonon mediated electron-electron interaction. (c) Series of diagrams corresponding to the RPA for the effective interaction in the presence of both electron-electron and electron-phonon interactions. The wiggly (dashed) line represents the electron-electron Coulomb (SO phonon mediated) interaction  and $\Pi_0$ the bare polarizability.
}
\label{fig:1}	
\end{figure}

\section{Theory}
We consider the two component effective Hamiltonian for both monolayer and bilayer graphene, which is given by\cite{mccann2006, mccann2007}
\begin{equation}
H_J(\bm{k})=t_{\perp} \left(\frac{\hbar{v|{\bm k}|}}{t_{\perp}}\right)^J
\left(
\begin{array}{cc}
0 & e^{-i J\phi_{\bm {k}}} \\
e^{i J\phi_{\bm {k}}} & 0 \\
\end{array}
\right),
\end{equation}
where $J\!=\!1(J\!=\!2)$ corresponds to monolayer (bilayer) graphene, $t_{\perp}$ is the nearest interlayer hopping,  $v$ is the in-plane velocity of monolayer graphene, $|{\bm k}|=\sqrt{k_x^2+k_y^2}$ and $\phi_{\bm{k}}=\tan^{-1}(k_y/k_x)$. The corresponding energy eigenvalues and eigenfunctions are given by $\varepsilon_{\bm k,s}=st_{\perp}(\hbar v|{\bm k}|/t_{\perp})^J$ and $\left|s,\bm{k}\right>={1 \over \sqrt{2}}\left(s,e^{i
  J\phi_{\bm{k}}}\right)$ where $s=\pm 1$ are band indices.

For monolayer graphene, this model holds over the Fermi energy range limited by intralayer coupling ($\gamma_0\sim$3 eV) while for bilayer graphene it is valid for the Fermi energy range limited by the interlayer coupling ($\gamma_1\sim$0.4 eV), which is still within the range we are considering in this work. Note that if the Fermi energy or the corresponding carrier density is high enough, the interlayer hopping $t_{\perp}$ can be negligible and the bilayer graphene is simply described by a collection of two monolayer graphene sheets. We will discuss the high carrier density limit in bilayer graphene later.

In the coupled system, electrons interact with each other through the direct Coulomb interaction $v_c(q)=2\pi  e^2/\epsilon_{\infty}q$ [Fig.~\ref{fig:1}(a)] and the SO-phonon-mediated interaction $v_{\rm ph}(q,\omega)=M_0^2(q) D_0(\omega)$ [Fig.~\ref{fig:1}(b)]. Here the electron-SO phonon coupling $M_0(q)$ is given by\cite{mahan}
\begin{equation}
M_0^2(q)=v_c(q) \alpha e^{-2qd}{\omega_{\rm SO} \over 2},
\end{equation}
where $d$ is the distance between the graphene and the polar substrate,
\begin{equation}
\alpha = \epsilon_{\infty} \left [ \frac{1}{\epsilon_{\infty}+1} - \frac{1}{\epsilon_0 +1} \right ],
\end{equation}
$\epsilon_0$ ($\epsilon_{\infty}$) is the zero- (high) frequency dielectric constant, and $D_0(\omega)$ is the bare SO-phonon propagator given by
\begin{equation}
D_0(\omega)={2\omega_{\rm SO} \over \omega^2-\omega_{\rm SO}^2}.
\end{equation}
Within the RPA the effective electron-electron interaction is obtained from the sum of all bare bubble diagrams [Fig.~\ref{fig:1}(c)] and is given by
\begin{equation}
v_{\rm eff}(q,\omega)=\frac{v_c(q) + v_{\rm ph}(q,\omega)}{1 - \left [ v_c(q)
+ v_{\rm ph}(q,\omega) \right ] \Pi_0(q,\omega)}
=\frac{v_c(q)}{\epsilon_{\rm t}(q,\omega)},
\label{eq:v_eff}
\end{equation}
where $\Pi_0(q,\omega)$ is the bare polarizability of graphene. 
Thus the total dielectric function from electrons and SO phonons is given by\cite{ph_el}
\begin{align}
\epsilon_{\rm t}(q,\omega) &=  
 1- v_c(q)
\Pi_0(q,\omega) +
\frac{M_0^2(q)D_0(\omega)}{v_c(q)+M_0^2(q)D_0(\omega)}\nonumber\\
 &= 1- \frac{2 \pi e^2}{\epsilon_{\infty}q}
\Pi_0(q,\omega) 
\frac{\alpha e^{-2qd}}{1 - \alpha e^{-2qd} -
\omega^2/\omega_{\rm SO}^2}.
\end{align}
Here for simplicity we use the zero-temperature polarizability as an approximation, which is valid in monolayer graphene at typical doping densities $n=10^{11}$--$10^{13}$ cm$^{-2}$ because the corresponding Fermi temperature $T_{\rm F}=400$--$4000$ K is much larger than room temperature. On the other hand, in bilayer graphene the Fermi temperature at low densities $n\sim10^{11}$ cm$^{-2}$ ($T_{\rm F}\sim40$ K) is smaller than room temperature, and thus the zero temperature approximation is valid only at relatively high densities $n>10^{12}$ cm$^{-2}$($T_{\rm F}\sim400$ K) in bilayer graphene.

Alternatively, the effective interaction $v_{\rm eff}(q,\omega)$ can be written as the sum of the screened electron-electron Coulomb interaction and the screened electron-SO phonon interaction, \cite{mahan,2DEG}
\begin{equation}
v_{\rm eff}(q,\omega)=\frac{v_c(q)}{\epsilon(q,\omega)}+v^{\rm sc}_{\rm ph}(q,\omega),
\label{eq:veff_sum}
\end{equation}
where 
$\epsilon(q,\omega)$=$1-v_c(q)\Pi_0(q,\omega)$ is the electronic dielectric function within the RPA.\cite{diel}
The screened electron-SO phonon interaction is given by $v^{\rm sc}_{\rm ph}(q,\omega)=[M(q,\omega)]^2D(q,\omega)$ where ${M}(q,\omega)=M_0(q)/\epsilon(q,\omega)$ is the screened interaction matrix element and $D(q,\omega)$ is the renormalized phonon propagator given by
\begin{align}
D(q,\omega)=\frac{D_0(\omega)}{1-[M_0(q)]^2D_0(\omega)\Pi_0(q,\omega)/\epsilon(q,\omega)}.
\label{eq:phonon_propagator}
\end{align}
Note that the interaction between electrons and SO-phonons is dynamically screened. The effect of dynamic screening in contrast to that of the static screening will be discussed later. 

The self-energy of the coupled system within the G$_0$W approximation is given by$\cite{inlife}$
\begin{align}
{\mathrm{Im}}\![\Sigma_s^{\mathrm{t}} &({\bm k},\omega)]\!=\!
\sum_{s'}\!\int\!\frac{d^2 q}{(2\pi)^2} \left [\Theta(\omega-\xi_{\bm k+\bm q,s'}) - \Theta(-\xi_{\bm k+\bm q,s'}) \right ] \nonumber \\ 
&\times\, \,{\rm Im}\!\left[v_{\rm eff}(q,\omega)\right] F_{ss'}({\bm 
  k},{\bm k}+{\bm q}),
\label{eq:self_energy}
\end{align}
where $F_{ss'}({\bm k},{\bm k}+{\bm q})=\frac{1}{2}(1+ss'\cos{J\theta_{\bm{k,k+q}}})$ is the wavefunction overlap factor and ${\theta_{\bm{k,k+q}}}$ is the angle between $\bm{k}$ and $\bm{k+q}$. Within the on-shell approximation, $\omega\ $is substituted by the on-shell energy  $\xi_{\bm{k},s} = \varepsilon_{\bm{k},s} - \mu$ where $\mu$ is the chemical potential. The scattering rate is given by the imaginary part of the self-energy via the relation $\hbar/\tau = -2\mathrm{Im}[\Sigma^{\rm t}_s]$.




\section{Results}
\subsection{Scattering rate}

\begin{figure}
\centering
	\includegraphics[scale=0.47]{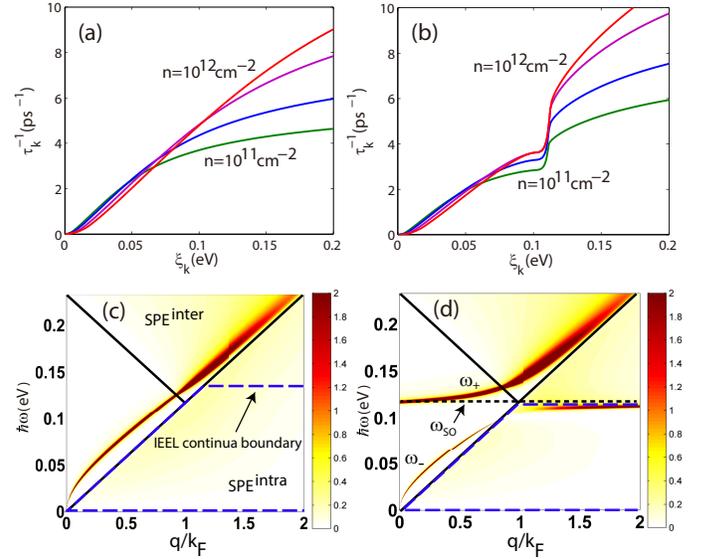}		
\caption{\label{fig:wide}(Color online) (a) Calculated scattering rate as a function of the on-shell energy $\xi_{\bm{k}}$ for different carrier densities $n=(1,2,5,10) \times 10^{11}$ cm$^{-2}$ for (a) uncoupled and (b) coupled monolayer graphene ($J=1$), and calculated energy-loss function for (c) uncoupled and (d) coupled monolayer graphene at $n=10^{12}$  cm$^{-2}$. The dotted horizontal line in (d) represents the SO phonon frequency, and the dashed line represents the boundary of the IEEL continua for an electron injected with the energy 140 meV (c) and 106 meV (d). SPE$^{\rm intra}$ (SPE$^{\rm inter}$) represents the single-particle excitation region for the intraband (interband) electron-hole excitations. Note that for the coupled system the plasmon dispersion is partly covered by the IEEL continua and thus a decay process via plasmon emission is available.
}
\label{fig:2}	
\end{figure}

\begin{figure}
\centering
	\includegraphics[scale=0.45]{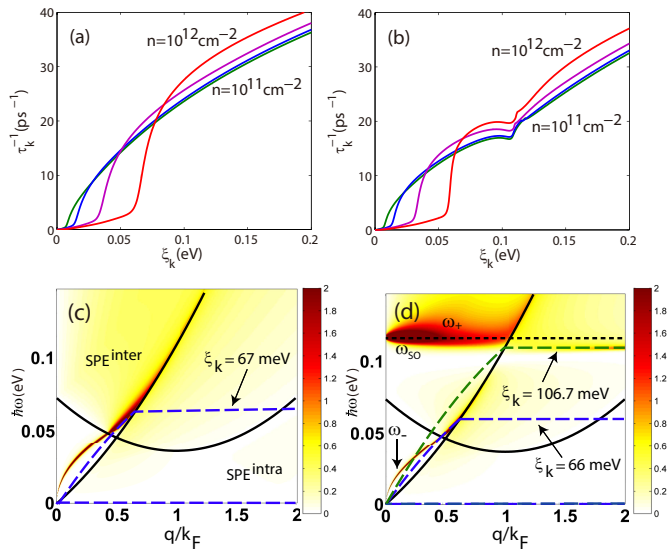}		
\caption{\label{fig:wide}(Color online) Calculated scattering rate as a function of $\xi_{\bm{k}}$ for different carrier densities $n=(1,2,5,10 )\times 10^{11}$ cm$^{-2}$ for (a) uncoupled and (b) coupled bilayer graphene ($J=2$),
and calculated energy loss function for (c) uncoupled and (d) coupled bilayer graphene  at $n=10^{12}$  cm$^{-2}$. In (d), the IEEL continua are drawn for two different energies of an injected electron: $\xi_{\bm{k}}=66$ meV  (blue dashed line) and $\xi_{\bm{k}}=106.7$ meV (green dashed line). At these energies, the scattering rate increases sharply because of the decay via the emission of the plasmonlike mode and phononlike mode, respectively, as shown in (b).
}
\label{fig:3}	
\end{figure}

For numerical calculations, the parameters\cite{param} corresponding to SiC are used throughout this paper:  $\hbar\omega_{\rm SO}=116.7$ meV, $\epsilon_{\infty}=6.4$, $\epsilon_{0}=10.0$, and $d=5$ $\rm \AA$. 
Figures \ref{fig:2}(a) and (b) show the scattering rates for uncoupled and coupled monolayer graphene as a function of the on-shell energy $ \xi_{\bm k}$. As in the case of the 2D electron gas with the parabolic energy dispersion, the scattering rate vanishes at the Fermi energy (i.e., $\xi_{\bm k}=0$) and shows the well-known quadratic behavior of $\sim\vert \xi_{\bm k}\vert^2 \ln\vert\xi_{\bm k}\vert$ near the Fermi energy. Away from the Fermi energy, an upward kink structure appears in the scattering rate for the coupled system, whereas the uncoupled system shows no such structure at any energy. 
Figures \ref{fig:2}(c) and 2(d) show the calculated loss functions --Im$[1/\epsilon_{\rm t}(q,\omega)]$ of the uncoupled and coupled monolayer graphene for carrier density $n=10^{12}$ cm$^{-2}$ along with the single-particle excitation (SPE) and the injected-electron energy loss (IEEL) continua. The loss function describes electronic energy dissipation and its poles represent the dissipation via plasmon excitations.
The intersections of the IEEL continua with the SPE continua indicate the allowed quasiparticle decay via electron-hole pair excitations while the intersections of the IEEL continua with the plasmon lines indicate the allowed quasiparticle decay via the emission of plasmons. 
Note that the volume of the IEEL continua depends on the energy with which an electron is injected. Hence, an injected electron with a higher energy can have more routes to decay than that injected with a low energy.

For the uncoupled monolayer graphene, the quasiparticle cannot decay via plasmon emission because the plasmon dispersion does not enter the IEEL continua over the whole energy range, as shown in Fig.~\ref{fig:2}(c). Thus the scattering rate in doped graphene does not show an abrupt increase at any energy.\cite{inlife} On the other hand, as shown in Fig.~\ref{fig:2}(d), for the coupled system there are two modes: the phononlike mode $\omega_+$  and the plasmonlike mode $\omega_-$.
The plasmonlike mode $w_-$ enters the IEEL continua at a finite critical wave vector $q_c\approx \omega_{\rm SO}(1-\alpha)/v$.\cite{ph_el} Thus an additional decay channel via $\omega_-$ emission is turned on around the SO phonon energy $E_{\rm SO}=\hbar\omega_{\rm SO}$, leading to an upward step in the scattering rate. Note that decay via the emission of the phononlike mode $\omega_{+}$ is impossible for monolayer graphene 
because the $\omega_{+}$ mode increases linearly at large $q$ with energy slightly higher than that of the uncoupled plasmon mode and thus it does not enter the IEEL continua. 

Figure \ref{fig:3} shows the scattering rate and the loss function in uncoupled and coupled bilayer graphene. Unlike the situation with monolayer graphene, the plasmon dispersions for both uncoupled and coupled bilayer graphene enter the IEEL continua. Therefore, even in the uncoupled system the scattering rate exhibits an abrupt increase, which is absent in the uncoupled monolayer system. In addition, while the coupled monolayer graphene system has only a single jump in the scattering rate, the coupled bilayer graphene shows two abrupt jumps. One of them occurs near $\xi_{\bm{k}} \approx E_{\rm SO}$, for which the emission of the phononlike mode  $\omega_{+}$ damped in the interband SPE is responsible. The other jump occurs due to the emission of the plasmonlike mode $\omega_{-}$ and its threshold energy strongly depends upon the carrier density, hence the emission of the plasmonlike mode $\omega_{-}$ is tunable with a carrier density. At higher carrier densities ($E_{\rm F} > E_{\rm SO})$ the plasmon is strongly coupled to the SO phonon and the threshold energy for the phononlike mode $\omega_{+}$ becomes a tunable quantity while the threshold energy for the plasmonlike $\omega_{-}$ is fixed around the SO phonon energy $E_{\rm SO}$.\cite{ph_el} Note that in bilayer graphene the step of the scattering rate at the phononlike mode shows a weaker density dependence on the carrier density than that in monolayer graphene because of the constant density of states. Also note that our calculation for bilayer graphene is based on the two band model which is valid only at low carrier densities. At high enough carrier densities, the interlayer coupling becomes negligible and the energy band structure of bilayer graphene behaves as a collection of monolayer graphene sheets, thus we expect that the results for bilayer graphene are similar to those of monolayer graphene. 
At such high densities, the density-dependent jump would disappear and only one single jump around $E_{\rm SO}$ would be found in the scattering rate, as in the monolayer case. We also note that high-energy plasmon modes are known to exist in bilayer graphene.\cite{Sensarma2010,Gamayun2011, Borghi2009} When an electron is injected with sufficiently high energies, a decay through the emission of the high-energy plasmon modes could be possible, leading to additional jumps in the scattering rate. Such jumps are not captured by the two-band effective model and thus beyond the scope of this paper.



\subsection{Mean free path}

\begin{figure}
\centering
	\includegraphics[scale=0.49]{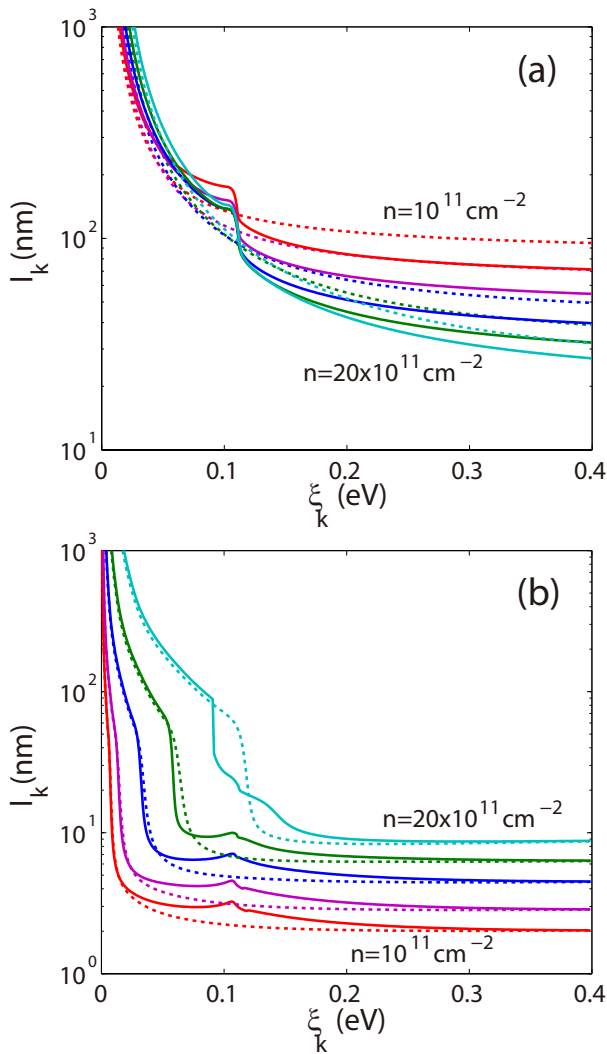}		
\caption{\label{fig:wide}(Color online) Calculated mean free path as a function of $\xi_{\bm{k}}$ for different carrier densities $n=(1,2,5,10,20) \times 10^{11}$ cm$^{-2}$ for (a) monolayer graphene and (b) bilayer graphene. The solid (dotted) lines represent the mean free paths in the presence (absence) of the electron-SO-phonon coupling. }
\label{fig:4}	
\end{figure}

The inelastic mean free paths $l_k=v_k \tau_k$ of the uncoupled and coupled monolayer graphene systems are given in Fig.~\ref{fig:4}(a). Unlike the uncoupled system, the coupled system shows a sharp decrease in the inelastic mean free path at the SO phonon energy $E_{\rm SO}$ at which the emission of plasmonlike mode is turned on and an electron loses its energy significantly.
Thus, by injecting an electron below or above $E_{\rm SO}$, we can change the mean free path of the system significantly, which can be used for designing a lateral hot electron transistor device in the coupled monolayer graphene system.\cite{Palevski1989,Sakamoto2000,BL} 
Figure \ref{fig:4}(b) shows the inelastic mean free path for bilayer graphene systems. 
The mean free path in bilayer graphene is much shorter than that in monolayer graphene and the step of the mean free path is very small around $E_{\rm SO}$. Thus the bilayer graphene may not be  a good candidate for the application to a lateral hot electron transistor utilizing the electron-phonon interaction.
At low energies, however, the inelastic mean free 
path strongly depends on the carrier density because the threshold energy of the plasmonlike mode $w_{-}$ is approximately proportional to the Fermi energy. Thus, by changing the carrier density, we can activate or deactivate the decay process via emission of the plasmonlike mode $w_{-}$, which is possible even in the uncoupled system without SO phonons. With the help of density tunability through gating, we can achieve a significant change in the mean free path. We find that the mean free path of an electron with an energy $\xi_{\bm{k}}\approx0.1$ eV  at the carrier density $n\sim10^{12}$ cm$^{-2}$ is $l\sim10^2$ nm but $l\sim$1--10 nm at $n\sim10^{11}$ cm$^{-2}$. This promises a possible use of bilayer graphene as a lateral-hot electron transistor in the absence of electron-SO-phonon coupling. 


\section{Discussion and Summary}
\begin{figure}
\centering
	\includegraphics[scale=0.34]{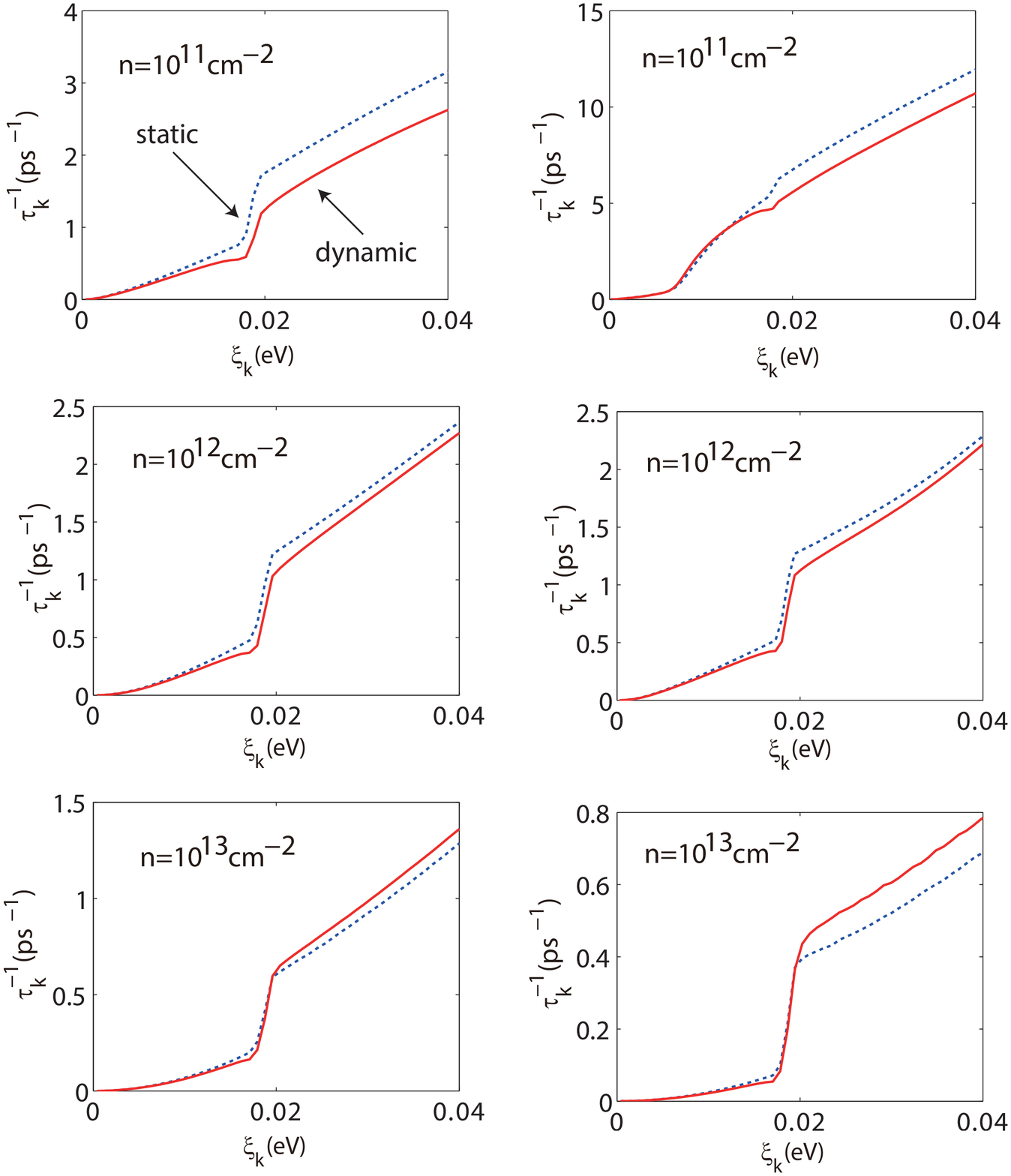}		
\caption{\label{fig:wide} Calculated scattering rate as a function of
  $\xi_{\bm{k}}$ for different carrier densities $n=10^{11}, 10^{12},
  10^{13}$ cm$^{-2}$ for monolayer graphene (left column) and for
  bilayer graphene (right column). The solid (dotted) lines represent
  the scattering rates with dynamically (statically) screened
  electron-SO-phonon interaction. Here $\hbar\omega_{\rm SO}=20$ meV
  is used for this calculation.} 
\label{fig:5}	
\end{figure}

In the previous section we showed the self-energy calculated with dynamical
RPA screening. In this section we discuss the effect of dynamical screening of the
electron-SO-phonon interaction compared with the results calculated with static
screening. We also discuss the effects of finite SO-phonon lifetime and multiple SO-phonon modes of the polar substrate.  

The static screening is equivalent to
putting $\omega=0$ in $\Pi_0 (q,\omega)$ in Eq.~(\ref{eq:phonon_propagator}), 
which means that the electrons screen the SO phonons statically,
while electron-electron interactions are treated dynamically to take
into account the plasmon effects.  Figure \ref{fig:5} shows the scattering rates for different carrier densities in monolayer and bilayer graphene with the static or dynamic screening approximations. We find that at low densities the scattering rates with static screening are larger than those with dynamic screening for both monolayer and bilayer graphene, while at high densities the results with dynamic screening are larger than those with static screening.
These effects may arise from 
the attractive nature of the phonon-mediated effective interaction, which originates from the retardation effect coming from the mass difference between ions and electrons. At low densities, the negative contribution to the scattering rate from the attractive effective interaction is significant, and the static screening suppresses the negative contribution more strongly than the dynamic screening, giving a larger scattering rate. As the carrier density increases, the negative contribution to the scattering rate becomes negligible, and the scattering rate with the static screening approximation becomes smaller than that with the dynamic screening, as shown in the bottom panel of \ref{fig:5}.

In our calculations, we assume an infinite lifetime for SO phonons because the inclusion of the finite lifetime just suppresses the jumps in the scattering rate without making any qualitative change. Furthermore, the typical decay rates of SO phonons are so small ($\sim$1 meV) compared to SO phonon energies that the effect of the finite phonon lifetime is quantitatively negligible.


We also assume in our calculations that only one single SO phonon mode is present in substrates. 
If multiple SO phonon modes exist, as in SiO$_2$, each mode couples to the electron motion independently, giving a jump in the scattering rate near the phonon mode energy.  Our results based on one single phonon mode correspond to each of these jumps, and thus extension to multiple phonon modes from our results is trivial. 




In conclusion, we theoretically calculated the inelastic scattering rates $\tau^{-1}$ and the hot-electron inelastic mean free paths $l$ for both monolayer and bilayer graphene on a substrate made of polar materials, treating the electron-electron interaction and the electron-SO-phonon interaction on an equal footing. 
In our theoretical calculation, we include the interaction between electrons, SO phonons, and plasmons within the RPA (for dynamical screening and phonon self-energy correction) and within the leading-order self-energy in the effective total interaction.
We find that the strong coupling between the SO phonon and the plasmon leads to an additional decay channel for the quasiparticles through the emission of the coupled mode and gives rise to an abrupt increase in the scattering rate, which is absent in the uncoupled system. In monolayer graphene a single jump in the scattering rate occurs around $E_{\rm SO}$, arising from the emission of the low energy branch of the coupled modes. By varying the energy of an injected electron, we can change the mean free path significantly. In bilayer graphene the emission of both low and high energy branches of the coupled modes contributes to the scattering rate, and gives rise to two abrupt changes in the scattering rate. 
In particular, in bilayer graphene, the emission of the plasmonlike mode  depends strongly on the carrier density while the threshold energy for the emission of the phononlike mode is weakly dependent on the carrier density and fixed at $E_{\rm SO}$. Utilizing the density dependence of  the plasmonlike mode, we can achieve a significant difference in the mean free path by varying the carrier density.
With the help of the abrupt changes in $\tau^{-1}$ and $l$ near $E_{\rm SO}$, our results for both monolayer and bilayer graphene are applicable to making an electronic device by varying the energy of an injected electron near $E_{\rm SO}$ or tuning the carrier density. 
It may be possible to fabricate a hot-electron transistor device with a sudden onset of negative differential resistance associated with sharp changes in the inelastic mean free path due to electron-coupled-mode scattering of the injected electrons. 

\acknowledgments 
This research was supported by the Basic Science Research Program through the National Research Foundation of Korea (NRF) funded by the Ministry of Science, ICT and Future Planning (Grant No. 2012R1A1A1013963 and Basic Science Research Program 2009-0083540).


\begin{thebibliography}{34}


\bibitem{sarmarmp2011} S. Das Sarma, S. Adam, E. H. Hwang, and E. Rossi, Rev. Mod.
Phys. {\bf 83}, 407 (2011).

\bibitem{netormp2009} A. H. Castro Neto, F. Guinea, N. M. R. Peres, K. S. Novoselov,
and A. K. Geim, Rev. Mod. Phys. {\bf 81}, 109 (2009).

\bibitem{Wu2011}
Y. Wu, Y. Lin, A. a Bol, K. a Jenkins, F. Xia, D. B. Farmer, Y. Zhu, and P. Avouris, Nature {\bf 472}, 74 (2011).



\bibitem{chennatnano2008} 
J. H. Chen, C. Jang, S. Xiao, M. Ishigami, and M. S. Fuhrer, Nat. Nanotechnol. {\bf 3}, 206 (2008).

\bibitem{Dorgan2010}
V. E. Dorgan,M.-H. Bae, and E. Pop, Appl. Phys. Lett. {\bf 97}, 082112 (2010).

 \bibitem{Fei2011}
Z. Fei, G. Andreev, W. Bao, and L. Zhang, Nano Lett. {\bf 11}, 4701 (2011).


\bibitem{Robinson2009}
J. Robinson {\it et al}., Nano Lett. {\bf 9}, 2873 (2009).

\bibitem{Sutter2009}
P. Sutter, Nat.Mater. {\bf 8}, 171 (2009).

\bibitem{Liu2010}
Y. Liu and R. F. Willis, Phys. Rev. B {\bf 81}, 081406 (2010).

\bibitem{Koch2010} 
 R. J. Koch, T. Seyller, and J. A. Schaefer, Phys. Rev. B {\bf 82}, 201413 (2010).


\bibitem{dean2010}
C. R. Dean \textit{et al}., Nat. Nanotechnol. {\bf 5}, 722 (2010).

\bibitem{Ulstrup}
S. Ulstrup, M. Bianchi, R. Hatch, D. Guan, A. Baraldi, D. Alfè, L. Hornekær, and P. Hofmann, Phys. Rev. B {\bf 86}, 161402 (2012).

\bibitem{wang2013}
L. Wang, I. Meric, P. Y. Huang, Q. Gao, Y. Gao, H. Tran, T. Taniguchi, K. Watanabe, L. M. Campos, D. a Muller, J. Guo, P. Kim, J. Hone, K. L. Shepard, and C. R. Dean, Science {\bf 342}, 614 (2013).

\bibitem{Principi}
A. Principi, M. Carrega, M. B. Lundeberg, A. Woessner, F. H. L. Koppens, G. Vignale, and M. Polini, Phys. Rev. B {\bf 90}, 165408 (2014).

\bibitem{Zou2010}
K. Zou, X. Hong, D. Keefer, and J. Zhu, Phys. Rev. Lett {\bf 105}, 126601 (2010).




\bibitem{andormp1982} T. Ando, A. B. Fowler, and F. Stern, Rev. Mod. Phys. {\bf 54}, 437 (1982).


\bibitem{park2014}
C.-H. Park, N. Bonini, T. Sohier, G. Samsonidze, B. Kozinsky, M. Calandra, F. Mauri, and N. Marzari, Nano Lett. {\bf 14}, 1113 (2014).

\bibitem{sohier2014}
 T. Sohier, M. Calandra, C.-H. Park, N. Bonini, N. Marzari, and F. Mauri, Phys. Rev. B {\bf 90}, 125414 (2014).

\bibitem{hwang2008} 
E. H. Hwang and S. Das Sarma, Phys. Rev. B {\bf 77}, 115449 (2008). 

\bibitem{Zhang2014} 
S. H. Zhang, W. Xu, F. M. Peeters, and S. M. Badalyan, Phys. Rev. B {\bf 89}, 195409 (2014). 


\bibitem{min2011} 
Hongki Min, E. H. Hwang, and S. Das Sarma, Phys. Rev. B {\bf 83}, 161404(R) (2011).

\bibitem{efetov2010} 
Dmitri K. Efetov and Philip Kim, Phys. Rev. Lett. {\bf 105}, 256805 (2010).

\bibitem{Pachoud2010}
A. Pachoud, M. Jaiswal, P. K. Ang, K. P. Loh, and B. \"{O}zyilmaz,
Europhys. Lett. {\bf 92}, 27001 (2010).

\bibitem{dawlaty}
J. M. Dawlaty, S. Shivaraman, M. Chandrashekhar, F. Rana, and M. G. Spencer, Appl. Phys. Lett. {\bf 92}, 042116 (2008).



\bibitem{Yan2013}
H. Yan, T. Low, W. Zhu, Y. Wu, M. Freitag, X. Li, F. Guinea, P. Avouris, and F. Xia, Nat. Photonics {\bf 7}, 394 (2013).


\bibitem{Mason1985}
 B. A. Mason and S. Das Sarma, Phys. Rev. B {\bf 31}, 5223 (1985).
 
  \bibitem{mahan} 
G.D. Mahan, {\it Many Particle Physics}, 3rd ed. (Kluwer/Plenum, New York, 2000).

\bibitem{Tse2007}
W.-K. Tse and S. Das Sarma, Phys. Rev. Lett. {\bf 99}, 236802 (2007).

\bibitem{LeBlanc2011}
J. P. F. LeBlanc, J. P. Carbotte, and E. J. Nicol, Phys. Rev. B {\bf 84}, 165448 (2011).

\bibitem{Carbotte2013}
J. P. Carbotte, J. P. F. LeBlanc, and P. E. C. Ashby, Phys. Rev. B {\bf 87}, 045405 (2013).





\bibitem {Brar2014}
V. W. Brar, M. S. Jang, M. Sherrott, S. Kim, J. J. Lopez, L. B. Kim, M. Choi, and H. Atwater, Nano Lett. {\bf 14}, 3876 (2014).


 \bibitem{Zhu2014} 
X. Zhu, W. Wang, W. Yan, M. B. Larsen, P. Bøggild, T. G. Pedersen, S. Xiao, J. Zi, and N. A. Mortensen, Nano Lett. {\bf 14}, 2907 (2014).




\bibitem{ph_el} 
E. H. Hwang, R. Sensarma, and S. Das Sarma, Phys. Rev. B {\bf 82}, 195406 (2010).


\bibitem{Jablan2011} 
M. Jablan, M. Solja\v ci\' c, and H. Buljan, Phys. Rev. B {\bf 83}, 161409 (2011).

\bibitem{Ong2012} 
Z.-Y. Ong and M. V Fischetti, Phys. Rev. B {\bf 86}, 165422 (2012).


\bibitem{SSC} 
E. H. Hwang and S. Das Sarma, Phys. Rev. B {\bf 87}, 115432 (2013).

\bibitem{scharf2013} 
B. Scharf, V. Perebeinos, J. Fabian, and I. \v{Z}uti\'c, Phys. Rev. B {\bf 88}, 125429 (2013).


\bibitem{fratini2008} 
S. Fratini and F. Guinea, Phys. Rev. B {\bf 77}, 195415 (2008).

\bibitem{Schiefele2008} 
J. Schiefele, F. Sols, and F. Guinea, Phys. Rev. B {\bf 85}, 195420 (2012).

\bibitem{Li2008} 
X. Li, E. A. Barry, J. M. Zavada, M. B. Nardelli, and K. W. Kim, Appl. Phys. Lett. {\bf 97}, 232105 (2010).

\bibitem{mccann2006}
E. McCann, Phys. Rev. B {\bf 74}, 161403 (2006).

\bibitem{mccann2007}
E. McCann, D. S. L. Abergel, and V. I. Fal’ko, Solid State Commun. {\bf 143}, 110 (2007).

\bibitem{2DEG} 
R. Jalabert and S. Das Sarma, Phys. Rev. B {\bf 40}, 9723 (1989).

\bibitem{diel} 
E. H. Hwang and S. Das Sarma, Phys. Rev. B {\bf 75}, 205418 (2007).


\bibitem{hwang1995} 
E. H. Hwang and S. Das Sarma, Phys. Rev. B {\bf 52}, R8668 (1995).

\bibitem{inlife} 
E. H. Hwang, B. Y.-K. Hu, and S. Das Sarma, Phys. Rev. B {\bf 76}, 115434 (2007).

\bibitem{param} 
H. Nienhaus, T. U. Kampen, and W. M\"{o}nch, Surf. Sci. {\bf 324}, L328 (1995).


\bibitem{Sensarma2010}
R. Sensarma, E. H. Hwang, and S. Das Sarma, Phys. Rev. B {\bf 82}, 195428 (2010).

\bibitem{Gamayun2011}
Gamayun, Phys. Rev. B {\bf 84}, 085112 (2011).

\bibitem{Borghi2009}
G. Borghi, M. Polini, R. Asgari, and A. H. MacDonald, Phys. Rev. B {\bf 80}, 241402 (2009).

\bibitem{Palevski1989}  
A. Palevski, M. Heiblum, C. P. Umbach, C. M. Knoedler, A. N. Broers, and R. H. Koch, Phys. Rev. Lett {\bf 62}, 1776 (1989).

\bibitem{Sakamoto2000}  
T. Sakamoto, H. Kawaura, T. Baba, and T. Iizuka, Appl. Phys. Lett. {\bf 76}, 2618 (2000).


\bibitem{BL} 
W.-K. Tse, E. H. Hwang, and S. Das Sarma, Appl. Phys. Lett. {\bf 93}, 023128 (2008).






\end{thebibliography}
\end{document}